\documentclass[12pt,tightenlines]{revtex4}

\usepackage{graphicx} 
\usepackage{bm} 
\usepackage{amsmath}
\usepackage{amssymb}

\def\Ket#1{\left|#1\right\rangle} 
\def\Bra#1{\left\langle#1\right|}
\def\KetBra#1#2{\Ket{#1}\!\Bra{#2}} 
\def\Proj#1{\KetBra{#1}{#1}}
\def\Eins{\bm{1}} 
\def\ie{i.\,e.\ }
\def\eg{e.\,g.\ }

\def\CCdots{\!\cdot\!\cdot\!\cdot}
\def\CLdots{.\!.\!.}  
\def\BQuad{\!\!\!\!}

\begin{document}

\title{Private entanglement over arbitrary distances, even using noisy
  apparatus} 
\author{Hans Aschauer}
\author{Hans J. Briegel}

\affiliation{Sektion Physik, Ludwig-Maximilians-Universit\"at,
  Theresienstr.\ 37, D-80333 M\"unchen, Germany}


\begin{abstract}
We give a security proof of quantum cryptography based entirely on
entanglement purification.  Our proof applies to all possible attacks
(individual and coherent). It implies the security of cryptographic
keys distributed with the help of entanglement-based quantum
repeaters. We prove the security of the obtained \emph{quantum
channel} which may not only be used for quantum key distribution, but
also for secure, albeit noisy, transmission of quantum information.
\end{abstract}
\pacs{PACS: 3.67.Dd, 3.67.Hk, 3.65.Bz} 
\maketitle


Quantum cryptography (QC) promises the security of data transmission
against any eavesdropping attack allowed by the laws of physics. The
first QC protocol was described by Bennett and Brassard as early as
1984 \cite{bb84}. Later, in 1991 Ekert presented a scheme based on
Bell's theorem \cite{ekert91}. Though the security of these protocols
is easy to prove under ideal conditions, a lot of work has been spent
to prove the security under realistic circumstances. In all QC
protocols, a possible eavesdropper is identified because of the
disturbance that he or she introduces when trying to gain information
about a quantum state that is transmitted.  The problem is that every
quantum channel introduces innocuous noise itself, which cannot, in
principle, be distinguished from noise introduced by an
eavesdropper. For that reason, a proof of unconditional security of QC
has to assume that all noise in the channel is due to the interference
of an eavesdropper.

Two different techniques have been developed to deal with these
difficulties: \emph{Classical privacy amplification} allows the
eavesdropper to have partial knowledge about the raw key built up
between the communicating parties Alice and Bob. From the raw key, a
shorter key is ``distilled'' about which Eve has vanishing (\ie
exponentially small in some chosen security parameter) knowledge.
Despite of the simple idea, proofs taking into account all
eavesdropping attacks allowed by the laws of quantum mechanics have
shown to be technically involved \cite{mayers,biham,hitoshi}.
Recently, Shor and Preskill \cite{shor} have given a simpler physical
proof relating the ideas in \cite{mayers,biham} to quantum error
correcting codes \cite{CSS,steane95}. \emph{Quantum privacy
  amplification} (QPA) \cite{deutsch96}, on the other hand, employs an
entanglement purification \cite{bennett96,bennett96a} protocol that
eliminates any entanglement with an eavesdropper by creating a few
perfect EPR pairs out of many imperfect (or impure) EPR pairs. In
principle, this method guarantees security against any eavesdropping
attack. However, the problem is that the QPA protocol assumes ideal
quantum operations.  In reality, these operations are themselves
subject to noise. As shown in \cite{briegel,duer_briegel,giedke},
there is an upper bound $F_{\text{max}}$ for the achievable fidelity
of EPR pairs which can be distilled using noisy apparatus. \emph{A
  priori}, there is no way to be sure that there is no residual
entanglement with an eavesdropper.  This problem could be solved if
Alice and Bob had fault tolerant quantum computers at their disposal,
which could then be used to reduce the noise of the apparatus to any
desired level. This was an essential assumption in the security proof
given by Lo and Chau \cite{lo}.

In this paper, we show that the standard two-way entanglement
purification protocols alone, with some minor modifications to
accomodate certain security aspects which will be discussed below, can
be used to efficiently establish a \emph{perfectly private quantum
  channel}, even when both the physical channel connecting the parties
and the local apparatus used by Alice and Bob are noisy.  This is of
particular interest because, as we show, the \emph{security} threshold
for the noise-level of the apparatus practically coincides with the
\emph{purification} threshold, so that the methods used for
long-distance quantum communication, using
entanglement-purification-based quantum repeaters
\cite{briegel,duer_briegel} can be used for secure quantum
communication \emph{without any further requirements}. In particular,
no fault tolerant quantum computers are required. This goal is
achieved by proving that the final state of the protocol factorizes
into a product state of the eavesdropper on one side, and Alice, Bob
and their laboratories (apparatuses) on the other side. Colloquially
speaking, we prove that Eve is factored out under the action of the
purification protocol, \ie the finite fidelity at the end of the
protocol is only due to entanglement with the apparatus.  Our proof
applies to all possible attacks (individual, collective, and coherent)
and can be utilized directly in long-distance quantum communication.
Different from existing work, we (i) prove the security of the entire
quantum channel, (ii) do not require fault tolerant quantum computers,
and (iii) our results have practical relevance, as the accuracy of the
apparatus used by Alice and Bob may be about two orders of magnitude
lower than the threshold accuracy for fault tolerant quantum computers
\cite{briegel,preskill97}.


The scenario is the following. Initially, Alice and Bob share a
numbered ensemble of $2N$ qubits $\{(a_1,b_1), \dots, (a_N,b_N)\}$,
$N$ qubits on each side, where $N$ is large.   Most generally,
the state they obtain can be written in the form
\begin{equation}
\rho_{AB}=\BQuad\sum_{\mu_1\dots\mu_N \atop \mu_1' \dots\mu_N'} \BQuad
\alpha_{\mu_1\CLdots\mu_N \atop \mu_1' \CLdots\mu_N'} |{\cal
B}_{\mu_1}^{(a_1b_1)}\CCdots {\cal B}_{\mu_N}^{(a_Nb_N)}\rangle \!
\langle {\cal B}_{\mu_1'}^{(a_1b_1)}\CCdots {\cal
B}_{\mu_N'}^{(a_Nb_N)}|
\label{rhoAB_1}
\end{equation}  
where $|\mathcal{B}_{\mu_j}^{(a_jb_j)}\rangle$, ${\mu_j}=00,01,10,11$
denote the 4 Bell states associated with the two particles $a_j$ and
$b_j$.  Specifically, $\Ket{\mathcal{B}_{00}} \equiv \Ket{\Phi^+} =
\left(\Ket{00}+\Ket{11}\right)/\sqrt{2}$, $\Ket{\mathcal{B}_{01}}
\equiv \Ket{\Psi^+} = \left(\Ket{01}+\Ket{01}\right)/\sqrt{2}$,
$\Ket{\mathcal{B}_{10}} \equiv \Ket{\Phi^-} =
\left(\Ket{00}-\Ket{11}\right)/\sqrt{2}$, $\Ket{\mathcal{B}_{11}}
\equiv \Ket{\Psi^-} = \left(\Ket{01}-\Ket{10}\right)/\sqrt{2}$. The
qubits have been distributed through some noisy channel, which may
also include repeater stations involving additional qubits
\cite{briegel}. In general, (\ref{rhoAB_1}) will be an entangled state
of $2N$ particles, which allows for the possibility of so-called
coherent attacks \cite{coll_attacks}. This state may be used to
establish a perfectly secret quantum channel, given that there is one
Bell state, say $|{\cal B}_{00}\rangle$, such that $\langle {\cal
B}_{00}^{(a_jb_j)}| \rho_{a_jb_j} |{\cal B}_{00}^{(a_jb_j)} \rangle >
F_{\text{min}}>1/2$ for the reduced density operator of every pair of
qubits $(a_j,b_j)$, where the exact value of $F_{\text{min}}$ depends
on the noise parameters of Alice's and Bob's apparatus \cite{briegel,
giedke}.

Upon reception of all pairs, Alice and Bob apply the following
protocol to them. Note that steps 1 and 2 are only applied once, while
steps 3, 4, and 5 are applied recursively. \newline %
{\tt Step 1:} On each pair of particles $(a_j,b_j)$, they apply
randomly one of the four bi-lateral Pauli rotations
$\sigma_k^{(a_j)}\otimes \sigma_k^{(b_j)}$, where k =
0,1,2,3. \newline %
{\tt Step 2:} Alice and Bob randomly renumber the pairs,
\((a_j,b_j)\to (a_{\pi(j)},b_{\pi(j)})\) where $\pi(j)$, $j=1,\dots,
N$ is a random permutation. \newline%
Steps 1 and 2 are required in order to treat correlated pairs
correctly. Note that steps 1 and 2 would also be required --- as
``preprocessing'' steps --- for the ideal protocol \cite{deutsch96},
if one requires that the protocol converges for arbitrary states of
the form (\ref{rhoAB_1}) to an ensemble of pure EPR states.  While in
\cite{deutsch96} it is possible to check whether or not the protocol
converges to the desired pure state, by measuring the fidelity of some
of the remaining pairs, this is not possible when imperfect apparatus
is used. Since the maximum attainable fidelity \(F_{\mathrm{max}}\) is
smaller than unity, there is no known way to exclude the possibility
that the non-ideal fidelity is due to correlations between the initial
pairs. In both steps Alice and Bob discard the information which of
the rotations and permutations, respectively, were chosen by their
random number generator.  Thus they deliberately loose some of the
information about the ensemble which is still available to Eve.  After
step 1, their knowledge about the state is summarized by the density
operator
\begin{equation}
\tilde\rho_{AB}=\BQuad \sum_{\mu_1\dots\mu_N}\BQuad
p_{\mu_1\dots\mu_N} |{\cal B}_{\mu_1}^{(a_1b_1)}\CCdots {\cal
B}_{\mu_N}^{(a_Nb_N)}\rangle \langle {\cal B}_{\mu_1}^{(a_1b_1)}
\CCdots {\cal B}_{\mu_N}^{(a_Nb_N)}|
\end{equation}    
which corresponds to a \emph{classically correlated ensemble} of pure
Bell states. Since the purification protocol that they are applying in
the following steps maps Bell states onto Bell states, it is
statistically consistent for Alice and Bob to assume after step 1 that
they are dealing with a (numbered) ensemble of pure Bell states, where
they have only limited knowledge about which Bell state a specific
pair is in.  The fact that the pairs are correlated means that the
order in which they appear in the numbered ensemble may have some
pattern, which may have been imposed by Eve or by the channel
itself. By applying step 2, Alice and Bob $(i)$ deliberately ignore
this pattern and $(ii)$ randomize the order in which the pairs are
used in the subsequent purification steps.  For all statistical
predictions made by Alice and Bob, they may consistently describe the
ensemble by the density operator \footnote{While, strictly speaking,
this equality holds only for $N\to\infty$, the subsequent arguments
also hold for the exact but more complicated form of
(\ref{rho_product}) for finite $N$.}
\begin{eqnarray}
\label{rho_product}
\tilde{\tilde\rho}_{AB} &=& \left(\sum_{\mu}p_{\mu} |{\cal
B}_{\mu}\rangle \langle {\cal B}_{\mu}|\right)^{\otimes N} \equiv
(\rho_{ab})^{\otimes N}
\end{eqnarray} 
in which the $p_{\mu}$ describe the probability with which each pair
is found in the Bell state $|{\cal B}_{\mu}\rangle$.  At this point,
Alice and Bob have to make sure that $p_{00}\equiv F> F_{\text{min}}$
for some minimum fidelity $F_{\text{min}}>1/2$, which they can do by
statistical tests on a certain fraction of the pairs.  Next, Alice and
Bob apply one of the standard purification protocols as described in
\cite{bennett96,deutsch96}. For simplicity, we concentrate on the
protocol given in \cite{deutsch96}; for other recurrence protocols, a
similar proof could be given \footnote{Note that for noisy local
  operations the hashing protocol, which requires Alice and Bob to
  apply a large number of CNOT operations in every distillation step,
  usually performs much worse that the recurrence protocols. The
  reason for this lies in the fact that the noise introduced with
  every CNOT operation accumulates and rapidly shatters the potential
  information that could ideally be gained from the parity
  measurement.}. The protocol uses these steps:
\newline%
{\tt Step 3:} Bi-lateral rotations \(1/2(\Eins^{(a)}-i\sigma_x^{(a)})
\otimes (\Eins^{(b)}+i\sigma^{(b)}_x)\) are applied to all pairs
\((a,b)\). \newline%
{\tt Step 4:} To all pairs of pairs a bi-lateral CNOT operation
(BCNOT) is applied. \newline%
{\tt Step 5:} The target pair of the BCNOT operation is measured on
both sides in $z$-direction. If the measurement results coincide, the
control pair is kept, otherwise it is discarded.

Since Alice and Bob use imperfect apparatus, it has been shown
\cite{briegel,giedke} that these protocols converge towards a
mixed-state ensemble $\rho_{ab}^{(\infty)}$ with a maximum attainable
fidelity $F_{\text{max}}<1$. If the fidelity of the local operations
is moderate, the value of $F_{\text{max}}$ could be quite low ($80\%$, as
an example).

In the following we will show that, despite of such a poor attainable
fidelity, Alice and Bob may happily proceed to apply the purification
protocol to establish a secure quantum channel \footnote{The fact that
  already the rotations used in steps 1 and 2 will be subject to noise
  is immaterial. As no measurements are performed, all such noise may
  be entirely attributed to the channel.}.  We show that, as $F \to
F_{\text{max}}$, the entanglement of the ensemble with the
eavesdropper is reduced exponentially fast with the number of
purification steps. In each step of the protocol, we assume that the
apparatus they use introduces errors described by the following map
\footnote{We implicitly assume that Alice and Bob can trust their
  apparatus. Similar as in the work of Mayers and Yao,
  quant-ph/9809039, Alice and Bob need \emph{not} trust the apparatus
  that produces the qubits. In the scenario considered in the present
  paper the quantum EPR source may be entirely under
  external control (e.\,g. repeater stations) and is thus not trusted
  at all. Alice and Bob merely check, before step~3 of the protocol,
  the fidelity of the ensemble.}
\begin{equation}
\rho_{AB} \rightarrow  \sum_{\mu,\nu=0}^{3}
f_{\mu\nu}\sigma_{\mu}^{(a)}\sigma_{\nu}^{(b)} \rho_{AB}\sigma_{\mu}^{(a)}
\sigma_{\nu}^{(b)}\,,
\label{white_noise_1}
\end{equation}
where $a$ and $b$ denote the qubits which are acted upon locally. The
$f_{\mu\nu}$ can be interpreted as the joint probability that the
Pauli rotations $\sigma_\mu$ and $\sigma_\nu$ are applied to qubits
$a$ and $b$, respectively. Eq.\ (\ref{white_noise_1})  includes, for an
appropriate choice of the coefficients $f_{\mu\nu}$, the one and 
two qubit depolarizing channel and combinations thereof, as studied in
\cite{briegel,duer_briegel}, but is more general.

It is possible to include the laboratories degrees of freedom in the
description. Noise of the form (\ref{white_noise_1}) can be attributed
to some interaction with the apparatus, which is described by a map
\begin{eqnarray}
|E\rangle_L|\psi\rangle_{AB} \rightarrow
\sum_{\mu,\nu=0}^3|e_{\mu\nu}\rangle_L \sigma_{\mu}^{(a)}
\sigma_{\nu}^{(b)} |\psi^{(ab)}\rangle.
\end{eqnarray}
This map explicitly accounts for the state of the apparatus before and
after the interaction. The states $|e_{\mu\nu}\rangle$ are pairwise
orthogonal and have the norm $ \langle e_{\mu\nu}|e_{\mu\nu}\rangle =
f_{\mu\nu}$. It is important to note that the laboratory degrees of
freedom $\Ket{e_{\mu\nu}}$ can, in principle, be identified in any
physical environment that generates noise of the form
(\ref{white_noise_1}), if the specific interaction Hamiltonian is
known.

For our purpose, however, the physical details of the environment are
of no concern, and we may replace the real process by the following
scenario, where both Alice and Bob have a ``little demon'' (L) in the
laboratory. For simplicity, we concentrate on the demon in Alices
laboratory only. Note that the generalisation to noise in both labs is
trivial.  Before every purification step, the demon applies randomly
one of the sixteen rotations $\sigma_{\mu}^{(a)} \otimes
\sigma_{\nu}^{(b)}$ to the qubits involved it this step, and keeps a
record of which rotations he chose.  For example, in the case of
uncorrelated white noise (depolarizing channel), it leaves each qubit
in its state ($\sigma_0\equiv I$) with some probability $f_0$, but
rotates its state by $\sigma_{j}$ with equal probabilities
$f_j=\frac{1-f_0}{3}$.

By doing this, the demon may accumulate a record of all errors in the
history of each qubit throughout the process. Instead of keeping track
of this growing list, he updates in each purifications step a single
\emph{flag} $\phi\equiv (ij)$ that is associated with each of the
pairs. The aim of the error flag is to keep the information required
for ``undoing'' the random rotations that occured in the history of
each pair. Note that, while this can be done trivially for unitary
networks, the situation is quite different with the QPA destillation
protocol, which includes measurements. For the proof, we show that
there exists a \emph{flag update function} -- as discussed below --
which enables the lab demon at any time of the protocol, to assign
each of the pairs to one out of four subensembles of the total
ensemble of pairs and to keep track of each of these subensembles
seperately.  Technically, the flag consists of two classical bits,
called the error phase bit $i$ and the error amplitude bit $j$. The
update is done in the following way: If a $\sigma_x$ ($\sigma_z$,
$\sigma_y$) error occurs, L inverts the error amplitude bit (error
phase bit, both error bits). Whenever Alice and Bob agree publicly to
keep a control pair $P_1$ (because of coinciding measurement outcomes
on the target pair $P_2$, see step 5 of the protocol), L calculates a
function of the error flags of $P_1$ and $P_2$ that updates the flag
of $P_1$. The function is shown in table~\ref{tab:error_flags}.  Note
that the error flag which belongs to a pair is, by construction, only
a function of the error records.  It is important to realize that,
what the lab demon is doing is \emph{not} quantum error correction, as
he is not applying any correction operation on the qubits in the
course of the entire protocol.  Instead of calculating the flags
during the run of the protocol, they could equally be calculated after
the protocol is finished.
\begin{table}
  \begin{tabular}[t]{|r|cccc|}
    \toprule 
    &(00)&(01)&(10)&(11) \\ 
    \colrule 
    (00)&(00)&(00)&(00)&(10) \\ 
    (01)&(00)&(01)&(11)&(00) \\ 
    (10)&(00)&(11)&(01)&(00) \\
    (11)&(10)&(00)&(00)&(00) \\ 
    \botrule
  \end{tabular}
  \caption
  {The value (phase error,amplitude error) of the updated error flag
  of a pair that is kept after a QPA step, given as a function of the
  error flags of $P_1$ and $P_2$ (left to right and top to bottom,
  respectively). }
  \label{tab:error_flags}
\end{table}

As mentioned earlier, at each purification step, the lab demon divides
the total ensemble into four subensembles $\rho_{\text{AB}} ^{(ij)}$
corresponding to the value $(ij)$ of the error flag. Initially, before
the QPA protocol starts, he assigns some random or fixed values to the
labels, while the subensembles are all in the same state.  That is,
the error flags and the states of the pairs are initially completely
uncorrelated. It is noteworthy that Bell diagonality of the states
$\rho_{\text{AB}}^{(ij)} = A^{(ij)}\Proj{\mathcal{B}_{00}}+
B^{(ij)}\Proj{\mathcal{B}_{11}}+ C^{(ij)}\Proj{\mathcal{B}_{01}}+
D^{(ij)}\Proj{\mathcal{B}_{10}}$ is preserved. This is due to the fact
that all operations in the protocol map Bell states onto Bell states.

In the following, we analyse the purification process in terms of
these four different subensembles $\rho_{\text{AB}}^{(ij)}$. In total,
we have to keep track of 16 coefficients that occur in the expansion
of each of the $\rho_{\text{AB}}^{(ij)}$ in the Bell basis.  These
coefficients after the $(n+1)$-th QPA step are functions of the
coefficients after the $n$-th QPA step:
\begin{equation}
  \label{eq:recurrence}
  \begin{split}
    A^{(00)}_{n} &\rightarrow
    A^{(00)}_{n+1}(A^{(00)}_n,A^{(01)}_n,\ldots,D^{(11)}_n),\\
    A^{(01)}_{n} &\rightarrow
    A^{(01)}_{n+1}(A^{(00)}_n,A^{(01)}_n,\ldots,D^{(11)}_n),\\
    &\,\,\,\vdots \\ D^{(11)}_{n} &\rightarrow
    D^{(11)}_{n+1}(A^{(00)}_n,A^{(01)}_n,\ldots,D^{(11)}_n).\\
  \end{split}
\end{equation}

The explicit form of the 16 recurrence relations~(\ref{eq:recurrence})
can be given, but they are rather lengthy. They imply
a reduced set of 4 recurrence relations for the quantities
$A_n=\sum_{ij}A_n^{(ij)}, \ldots, D_n=\sum_{ij}D_n^{(ij)}$ which
describe the evolution of the total ensemble under the purification
protocol. For $n\to\infty$, these quantities converge towards a
fixpoint $(A_\infty,B_\infty,C_\infty,D_\infty)$ where
$A_\infty=F_{\text{max}}$ is the maximal attainable fidelity
\cite{duer_briegel}. Different from the fidelity $F_n\equiv A_n$, we
define the \emph{conditional fidelity} $F_n^{\text{cond}}=A_n^{(00)}+
B_n^{(11)}+ C_n^{(01)}+D_n^{(10)}$. This is the fidelity of the
ensemble that Alice and Bob could attain, if the lab demon disclosed
the error flags (or, for that matter, only the history of the random
rotations, from which the flags can be calculated): Depending on the
error flag of a pair, Alice could then choose a local rotation that
transforms the pair into the Bell state $\Ket{\mathcal{B}_{00}}$ with
probability $F^{\text{cond}}$.

Evaluation of the recurrence relation yields that there are three
different regimes of noise parameters: In the high-noise regime (low
values of \(f_{00}\)), no purification is possible; the protocol
converges to completely depolarized pairs. In the low-noise regime
(high values of \(f_{00}\)), the protocol purifies \emph{and} the
conditional fidelity converges to unity: the protocol is in the
security regime. Between these two regimes, just above the
purification threshold, there exists a very narrow third regime: The
protocol purifies, while the conditional fidelity does \emph{not}
converge to unity. It is not known whether or not secure communication
is possible in this regime. For the depolarizing channel, for example,
the intermediate regime is contained in the interval \(f_0 \in
(0.8983,0.8988)\), while the security regime covers the entire
interval \(f_0 \in [0.8988,1]\). The security regime thus coincides,
for all practical purposes, with the purification regime, but it is
interesting to see that these regimes are not strictly identical.  It
shows that the process of factorization is, in the situation of
imperfect apparatus, not trivially connected to the process of
purification.  More details about these regimes will be published
elsewhere.
When the protocol is in the security regime, both the fidelity $F_n$
and the conditional fidelity $F^{\text{cond}}_n$ reach their
respective fixpoints exponentially fast with the same exponents (see
Fig.~\ref{fig:logFidelity}). 
From this it follows that there exists a polynomial
relation between the resources used in the purification process
(number of initial pairs) and the security parameter \(1 -
F^{\text{cond}}\). All results obtained from the evaluation of the
recurrence relations (\ref{eq:recurrence}) were also checked with the
help of Monte Carlo simulations, in which the QPA protocol was applied
to typical ensembles of Bell states.

\begin{figure}[t]
  \begin{center}
    \includegraphics[scale=0.8]{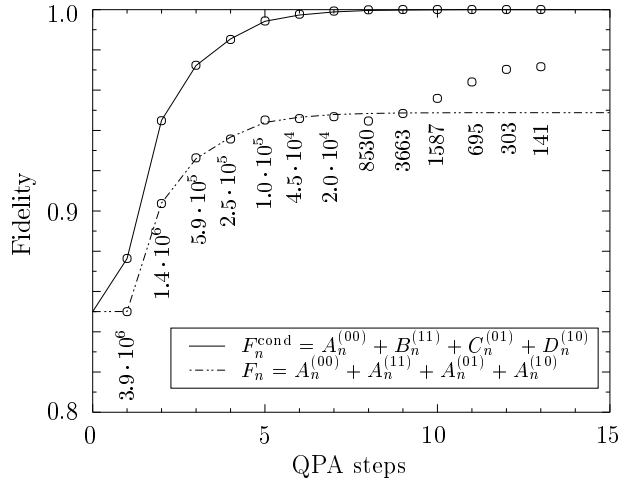}
    \caption{The fidelities $F$ and $F_{\text{cond}}$ as a function of the
      number of steps in the QPA protocol (analytical results (lines)
      and Monte Carlo simulation (circles)). For the calculation, one
      and two-qubit white noise with a noise fidelity of 97\% has been
      assumed. The Monte Carlo simulation was started with \(10^7\)
      pairs; the numbers indicate how many pairs are left after each
      step of the purification protocol. This decreasing number is the
      reason for the increasing fluctuations around the analytical
      curves.}
    \label{fig:Fidelity_data}
  \end{center}
\end{figure}

\begin{figure}[t]
  \begin{center}
    \includegraphics[scale=0.8]{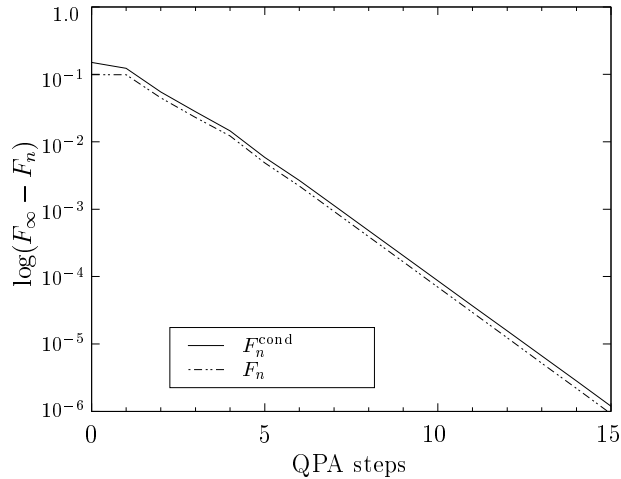}
    \caption{$F_\infty - F_n$ and $1 - F_n^{\text{cond}}$ plotted
      logarithmically against the purification step $n$. The
      parameters are the same as in
      Fig.~\protect{\ref{fig:Fidelity_data}}.}
    \label{fig:logFidelity}
  \end{center}
\end{figure}


Our results imply that the error flags and the states of the
subensembles become \emph{strictly correlated} during execution of the
purification protocol: The subensemble $(ij)$ ends in the state
$\Ket{\mathcal{B}_{ij}}$. In other words, the ``little demon'' has
acquired \emph{complete knowledge} about the states of all pairs after
sufficiently many purifications steps; the system consisting of the
pairs and the lab is thus in a pure state.  Now the same argument as
in \cite{deutsch96} applies: a system in a pure state cannot be
entangled with any other system --- any eavesdropper is factored
out, as his or her entanglement with the pairs is lost.

This proof can be extended to more general noise models if a slightly
modified protocol is used, where step 1 is repeated after every
distillation round \footnote{We are grateful to C.~H.~Bennett for
  pointing out this possibility.}. This effectively regularises any
type of local noise process to a process of the type
(\ref{white_noise_1}) that conserves the Bell diagonality of the
ensemble, for which we can apply the lab-demon interpretation
\footnote{Here, it is however required that Alice and Bob are able to
  perform one-qubit rotations used in step~1 well enough to keep the
  evolution Bell-diagonal.}.

The fact that the security regime of the protocol almost coincides with the
purification regime is of strong practical interest because it implies
that EPR pairs distributed over long distances with quantum repeaters
can be used for secure quantum communication without any additional
effort \footnote{The relevance of this result is underlined by a
recent proposal by Pan \emph{et al.} \cite{pan01}, who describe a
scheme of EPP using only optical elements, with which one would be able
to reach the desired error-threshold.}.

To summarize, Alice and Bob obtain, with the help of a standard
entanglement purification protocol, entangled EPR pairs. These pairs
have a limited fidelity $F\lesssim F_{\text{max}}<1$ which depends on
the noise introduced by local operations in their laboratory. Alice
and Bob may nevertheless use these pairs for secure quantum- or
classical communication, \eg teleportation \cite{bennett93} or key
distribution. At this stage, no further security tests are
necessary. Since we have shown that there exists no residual
entanglement with an eavesdropper, they may use all the pairs for the
key!  While there may be a significant error rate in the message,
Alice and Bob are allowed to apply classical error correction to the
transmitted message without disclosing any valuable information to
Eve.

\begin{acknowledgments}
We thank C. H. Bennett, A. Ekert, L. Hardy, H. Inamori, N. L\"utkenhaus,
R. Raussendorf, A. Schenzle and H. Weinfurter for valuable
discussions.  We are grateful to G. Giedke, N.  L\"utkenhaus, and
H.-K. Lo for constructive remarks on the manuscript.  This work has
been supported in part by the Schwerpunktsprogramm QIV of the DFG.
\end{acknowledgments}


\end{document}